\documentstyle[prl,aps,twocolumn,psfig]{revtex}
\newcommand{\ber}{\begin{eqnarray}}
\newcommand{\eer}{\end{eqnarray}}
\begin{document}
\twocolumn[\hsize\textwidth\columnwidth\hsize\csname
@twocolumnfalse\endcsname
\hfill 
\title{
Baryon Number Fluctuation and the Quark-Gluon Plasma
}
\author{
Zi-wei Lin and C.M. Ko
}
\address{
Cyclotron Institute and Physics Department,
Texas A\&M University, College Station, Texas 77843-3366
}
\maketitle

\begin{abstract}
We show that $\omega_B$ or $\omega_{\bar B}$, the squared baryon or 
antibaryon number fluctuation per baryon or antibaryon, is a possible 
signature for the quark-gluon plasma that is expected to be created 
in relativistic heavy ion collisions, as it is a factor of three smaller than
in an equilibrated hadronic matter due to the fractional baryon number 
of quarks. Using kinetic equations with exact baryon number conservation, 
we find that their values in an equilibrated matter are half of those
expected from a Poisson distribution. Effects due to finite acceptance 
and non-zero net baryon number are also studied. 
\end{abstract}
\pacs{25.75.-q, 12.38.Mh, 5.20.Dd, 05.40.-a}
]
\section{Introduction}

A new state of matter, the quark-gluon plasma, is 
expected to be formed in heavy ion collisions at ultra-relativistic energies, 
such as at the Relativistic Heavy Ion Collider (RHIC) that
has just begun its operation at the Brookhaven National Laboratory. 
Many observables have been proposed as possible signatures for the quark-gluon 
plasma phase during the collisions \cite{qms}, such as strangeness enhancement 
\cite{strange}, $J/\psi$ suppression \cite{jpsi}, modification of 
high $p_T$ particle spectrum \cite{quench}, and $M_T$ scaling \cite{mt} 
and double phi peaks \cite{double} in the dilepton spectrum. Recently, 
event-by-event fluctuations of various particles have also attracted 
much attention \cite{ebe}. Since the baryon and charge numbers of quarks 
are fractional, the net baryon number \cite{Yuki} and net charge \cite{Jeon} 
fluctuations per particle are smaller in the quark-gluon plasma than in the 
hadronic matter and have also been suggested as possible signatures for the 
quark-gluon plasma. 

Because of conservation of total net baryon number, 
the net baryon fluctuation is finite only if one considers 
a part of the full phase space, e.g., in a finite rapidity range.  
Since the quark-gluon plasma is eventually converted to the hadronic matter, 
elastic scatterings of baryons and antibaryons would modify their phase space 
correlations and thus increase the net baryon fluctuation towards that 
expected from an equilibrated hadronic matter \cite{Shuryak}. 
On the other hand, elastic scatterings of baryons and antibaryons 
do not change their numbers in the full phase space, so 
fluctuations of baryon and antibaryon numbers are 
alternative signatures for the quark-gluon plasma in
relativistic heavy ion collisions.

In this study, we shall use a master equation with exact baryon 
number conservation to derive the baryon and antibaryon number 
fluctuations in an equilibrated matter, and to study the effects due
to finite acceptance and non-zero net baryon number.

\section{Kinetic model}
\label{kinetic}

Due to the different fundamental units of baryon numbers in 
quark-gluon plasma and 
in hadronic matter, fluctuations of baryon and antibaryon 
numbers, like the fluctuation of the net baryon number, 
take very different values in the two phases of matter.  
To study dynamically the baryon number fluctuation in heavy ion
collisions, we introduce a kinetic model that takes into account
both production and annihilation of quark-antiquark or baryon-antibaryon 
pairs. We first consider the case of baryon-antibaryon production from 
and annihilation to two mesons, i.e., $m_1 m_2 \leftrightarrow B \bar B$, 
in a hadronic matter of net baryon number $s=0$ \cite{quark}.
Following the formalism of Ref.\cite{rate} for the production
of particles with conserving charges, we have the following
master equation for the multiplicity distribution of $B\bar B$ pairs:
\ber
\frac{dP_n}{d\tau}&=&{G\over V} \langle N_{m_1} \rangle 
\langle N_{m_2}\rangle\, \left (P_{n-1}-P_n \right ) \nonumber\\
&-& \frac {L}{V} \, \left [ n^2 P_n - (n+1)^2 P_{n+1} \right ].
\label{generaln}
\eer
In the above, $P_n(\tau)$ denotes the probability of finding $n$ pairs 
of $B \bar B$ at time $\tau$; $G \equiv \langle \sigma_G v \rangle$ and
$L \equiv \langle \sigma_L v \rangle$ are the momentum-averaged cross sections
for baryon production and annihilation, respectively; $N_k$ represents 
the total number of particle species $k$; and $V$ is the proper volume 
of the system.

The equilibrium solution to Eq.~(\ref{generaln}) is 
\ber
P_{n,\rm eq} &=& \frac {\epsilon^n} {I_0 ( 2\sqrt \epsilon )~(n!)^2},
\label{equil}
\eer
where $I_0$ is the modified Bessel function, and 
\ber
\epsilon\equiv \frac {G \langle N_{m_1} \rangle \langle N_{m_2}\rangle} {L}.
\label{eps}
\eer

Using the generating function at equilibrium, 
\ber
g_{\rm eq}(x) 
&\equiv& \sum_{n=0}^\infty P_{n, \rm eq} x^n 
=\frac {I_0 ( 2\sqrt {\epsilon x})}{I_0 ( 2\sqrt \epsilon)},
\eer
with $g(1) = \sum P_n = 1$ due to normalization of the multiplicity 
probability distribution, it is straightforward to obtain all moments 
of the equilibrium multiplicity distribution \cite{Cle0}. 
For example, the mean baryon number per event is given by 
\ber
\langle B \rangle_{\rm eq}=
b_0 \langle n \rangle=b_0 \frac {\partial g(1)}{\partial x_1}=
b_0 \sqrt \epsilon \frac {I_1}{I_0}
\simeq b_0 \sqrt \epsilon,  
\label{neq}
\eer
In the above, $b_0$ is the fundamental unit of baryon number in the matter, 
and the argument of Bessel functions $I_\nu$'s is $2\sqrt \epsilon$. 
In obtaining the last expression in Eq.~(\ref{neq}), 
we have kept only the leading term in $\sqrt\epsilon$, corresponding
to the grand canonical limit, $\sqrt \epsilon \gg 1$,
as baryons and antibaryons are abundantly produced in heavy ion
collisions at RHIC \cite{star,ampt}.   

The squared baryon number fluctuation is given by
\ber
D_B &\equiv& \langle B^2 \rangle-\langle B \rangle^2 
=b_0^2 \left [\frac {\partial g(1)}{\partial x_1}
+\frac {\partial^2 g(1)}{\partial x_1^2}
-\left (\frac {\partial g(1)}{\partial x_1}\right )^2
\right ] \nonumber \\
&=& b_0^2  \left [ \sqrt \epsilon \frac{I_1}{I_0} + \epsilon \left ( 
\frac{I_2}{I_0}-\frac{I_1^2}{I_0^2} \right )
\right ]
\simeq b_0^2 \frac {\sqrt \epsilon} {2}.
\label{db}
\eer

It is seen that the mean number of baryons 
is proportional to $b_0$ while their squared fluctuation is
proportional to $b_0^2$. The squared baryon number 
fluctuation per baryon is thus given by 
\ber 
\omega_{B,\rm eq}
&=& b_0 \left [ 1-\sqrt \epsilon \left ( 
{I_1 \over I_0}-{I_2 \over I_1} \right ) \right ]
\simeq {b_0 \over 2}.
\label{omegaeq}
\eer
The mean number of antibaryons $\langle\bar B\rangle$ and their fluctuation 
$\omega_{\bar B}$ are the same as those of baryons as a result of zero net
baryon number. Since $b_0$ is 1/3 in quark-gluon plasma and 1 in 
hadronic matter, $w_B$ and $w_{\bar B}$ are smaller in the 
quark-gluon plasma than in an equilibrated hadronic matter by a factor of $3$. 

We note that the equilibrium multiplicity distribution in Eq.~(\ref{equil}) 
is not Poisson, as pointed out earlier in Ref. \cite{Cle}. 
The non-Poisson distribution results from the quadratic dependence on the
multiplicity $n$ in the loss term of the master equation of 
Eq.~(\ref{generaln}) due to baryon number conservation.  
A Poisson distribution is obtained if the dependence on the multiplicity $n$ 
is linear, which corresponds to production of particles that do not carry
conserved charges.  The master equation of Eq.~(\ref{generaln}) 
also gives a Poisson distribution at early times ($\tau \rightarrow 0$) when 
the loss term can be neglected. This corresponds to either production 
of particles with conserved charges during the early stage of heavy ion 
collisions or particle production without chemical equilibration 
as in $e^+ e^-$ collisions. Since a Poisson multiplicity 
distribution gives
\ber
D_B^{\rm Poisson}=b_0 \langle B \rangle,
\eer
the squared baryon number fluctuation per baryon is
\ber
\omega_B^{\rm Poisson}=b_0,
\label{Poisson}
\eer
which is a factor of 2 larger than that in an equilibrated quark-gluon 
plasma or hadronic matter. A similar result has been obtained previously
by Gavin and Pruneau based on thermodynamic considerations \cite{Gavin}. 

\section{Finite Acceptance and Non-Zero Net Baryon}
\label{kinetic-accpt}

Because of experimental limitations, only protons and antiprotons 
in a certain rapidity and momentum range are usually measured. 
Moreover, the net baryon number in heavy ion collisions is 
in general non-zero even at mid-rapidity due to the presence of projectile 
and target nucleons. To generalize the master equation to include 
these effects, we first consider the case of two species of antibaryons,
e.g., antiproton and antineutron 
production from meson-meson interactions, i.e.,  $m_1 m_2 \leftrightarrow 
B \bar p$ and $m_1 m_2 \leftrightarrow B \bar n$ in a hadronic matter of 
net baryon number $s\geq 0$ \cite{quark}. Defining $P^{k,l}$ as the 
probability of finding $k$ number of antibaryon 
species $1$ and $l$ number of antibaryon species $2$ 
in an event ($k,l=0,\cdots,\infty$ for $s \geq 0$), 
we then have the following generalized master equation:
\ber
&&\frac{dP^{k,l}}{d\tau}=
{\epsilon_1 L_1 \over V} \left (P^{k-1,l}-P^{k,l} \right ) 
+{\epsilon_2 L_2 \over V} \left (P^{k,l-1}-P^{k,l} \right ) \nonumber\\
&&- \frac {L_1}{V} \, \left [ k (k+\!l+\!s) P^{k,l} 
- (k\!+\!1) (k\!+\!l+\!s+\!1) P^{k+1,l} 
\right ] \nonumber\\
&&- \frac {L_2}{V} \, \left [ l (k\!+\!l+\!s) P^{k,l} 
- (l\!+\!1) (k\!+\!l+\!s+\!1) P^{k,l+1} 
\right ].
\eer
In the above, $\epsilon_j$  ($j=1,2$) is defined as in Eq.~(\ref{eps}) 
for the process involving antibaryon species $j$, and $G_j$ and $L_j$  
denote the momentum-averaged cross sections for the gain and loss terms,  
respectively. We have neglected effects due to quantum statistics, as
they are not expected to be significant \cite{Bertsch}.

The equilibrium solution to the above equation is given by the product 
of the equilibrium solution for the total $B\bar B$ pair number $(k+l)$ 
and a binomial distribution, i.e., 
\ber
P_{\rm eq}^{k,l}
&=&\left [\frac {\epsilon^{k+l+s/2}} 
{I_s ( 2\!\sqrt \epsilon )(k\!+\!l)! (k\!+\!l\!+\!s)! } \right ]
\left [ \frac {(k\!+\!l)! f_1^k f_2^l} {k!\; l!} \right ],
\label{product}
\eer
with 
\ber
\epsilon &\equiv& \epsilon_1+\epsilon_2, ~f_j=\frac{\epsilon_j}{\epsilon}. 
\eer
The distribution in Eq.~(\ref{product}) can be understood intuitively 
as resulting from first obtaining the distribution for the total antibaryon 
number $(k+l)$, and then selecting $k$ number of antibaryon species $1$ 
and $l$ number of antibaryon species $2$ from the total antibaryons with 
probabilities of $f_1$ and $f_2$, respectively.

At equilibrium, the generating function $g (y_1,y_2) \equiv \sum  
P^{k,l} y_1^k y_2^l$ is given by:
\ber
g_{\rm eq} (y_1,y_2)
\!=\!\frac {\left (f_1 y_1\!\!+\!\! f_2 y_2 \right ) ^{-s/2}}
{I_s ( 2\sqrt \epsilon)} 
I_s \!\left ( \! 2\sqrt \epsilon 
\sqrt {f_1 y_1\!\! +\!\! f_2 y_2}\! \right )\!.
\label{equil2}
\eer

The above formulae can be equally applied to study the  
effects due to finite phase space coverage in experiments. In this case, 
antibaryon species 1 and 2 denote antibaryons in and out of the observed 
phase space, respectively; and $f_1$ and $f_2$ are their respective fractions. 

To generalize to $N$ numbers of baryon and antibaryon species, 
we consider $m_1 m_2 \leftrightarrow B_i \bar B_j$ ($i,j=1,\cdots,2N$),   
where $B_i$ and $\bar B_j$ with $i,j=1,\cdots,N$  
denote baryon species $i$ and antibaryon species $j$ in the observed
phase space, respectively, and $B_i$ or $\bar B_j$ with $i,j=N+1,\cdots,2N$ 
denote baryon species $(i\!-\!N)$ and antibaryon 
species $(j\!-\!N)$ out of the observed phase space, respectively.
The joint generating function for both baryons and antibaryons 
is defined by
\ber
&&g(x_1,\cdots,x_{2N};y_1,\cdots,y_{2N}) 
\equiv \nonumber \\
&&\sum P_{B_1,\cdots,B_{2N}}^{\bar B_1,\cdots,\bar B_{2N}}
\left ( \prod_{i=1}^{2N} x_i^{B_i} \right )
\left ( \prod_{j=1}^{2N} y_j^{\bar B_j} \right ),
\eer 
with the summation over all baryon and antibaryon numbers,  
$B_i$'s and $\bar B_j$'s ($i,j=1,\cdots,2N)$.  
Assuming that particle momentum distributions are thermal, 
the generating function at equilibrium is given by
\ber
&&g_{\rm eq}(x_1,\!\cdots\!,x_{2N};y_1,\!\cdots\!,y_{2N}) = 
\frac {\left ( \sum h_i x_i \right ) ^{s/2}
\left ( \sum f_j y_j \right ) ^{-s/2}}
{I_s \left ( 2\sqrt \epsilon \right )} \nonumber \\
&&\times I_s \left ( 2 \sqrt \epsilon \sqrt {\sum h_i x_i} 
\sqrt {\sum f_j y_j} \right ),
\label{equiln}
\eer
where
\ber
\epsilon \equiv \sum_{i,j} \epsilon_{ij},
\eer
with $\epsilon_{ij}$ defined for the process 
$m_1 m_2 \leftrightarrow B_i \bar B_j$ as in Eq.~(\ref{eps}), and 
\ber
 h_i=\left (\sum_j \epsilon_{i j} \right )/\epsilon,  
~f_j=\left (\sum_i \epsilon_{i j} \right )/\epsilon. 
\eer

From Eq.~(\ref{equiln}), we obtain the mean numbers of baryons
and antibaryons, 
\ber
\langle B_i \rangle_{\rm eq} 
&=&b_0 h_i \left (\! \sqrt \epsilon 
\frac {I_{s+1}}{I_s}\! +\! s\! \right )
\simeq b_0 h_i \sqrt \epsilon 
\left (\!1\!+\!\frac {2s\!-\!1}{4\sqrt \epsilon}\! \right ),
\label{beq}\\ 
\langle \bar B_j \rangle_{\rm eq} 
&=& b_0 f_j \sqrt \epsilon ~\frac {I_{s+1}}{I_s}
\simeq b_0 f_j \sqrt \epsilon \left (1-\frac {2s+1}{4\sqrt \epsilon} \right ), 
\label{bbeq}
\eer
and their fluctuations, 
\ber
\omega_{B_i,\rm eq} 
&=&b_0 \left [1- h_i \left ( \frac { \epsilon \left (
\frac {I_{s+1}^2}{I_s^2} -\frac {I_{s+2}}{I_s} \right )+s}
{\sqrt \epsilon \frac {I_{s+1}}{I_s}+s}\right ) \right ] \nonumber \\
&\simeq & b_0 \left [1- {h_i \over 2} 
\left (1+\frac {2s-1}{4\sqrt \epsilon} \right ) \right ], 
\label{finiteb} \\
\omega_{\bar B_j, \rm eq} 
&=&b_0 \left [ 1-f_j \sqrt \epsilon \left (
\frac {I_{s+1}}{I_s} -\frac {I_{s+2}}{I_{s+1}} \right ) \right ] \nonumber \\
&\simeq & b_0 \left [1- {f_j \over 2} 
\left (1-\frac {2s+1}{4\sqrt \epsilon} \right ) \right ].
\label{finitebb} 
\eer
In the above, 
we have kept only the first two leading terms in the expansions  
as $s/\sqrt \epsilon \ll 1$ for ultra-relativistic heavy ion collisions.  
We note that Eqs.~(\ref{beq}-\ref{bbeq}) and (\ref{finiteb}-\ref{finitebb}) 
reduce to Eqs.~(\ref{neq}) and (\ref{omegaeq}) in the special case of 
full phase space coverage with only one baryon and antibaryon species 
($h_1=f_1=1$) and $s=0$.  

Keeping only the leading terms in Eqs.~(\ref{finiteb}) and (\ref{finitebb}) 
leads to 
\ber
 \omega_{B_i, \rm eq} \simeq b_0 \left (1-{h_i \over 2} \right ),     
~\omega_{\bar B_j, \rm eq} \simeq b_0 \left (1-{f_j \over 2} \right ),
\label{finite} 
\eer
where $h_i$ and $f_j$ ($i,j=1,\cdots,N$) represent, respectively, 
the fraction of total baryons that are observed baryon species $i$ 
and the fraction of total antibaryons that are observed antibaryon species $j$.
Eq.(\ref{finite}) thus shows 
that finite acceptance introduces a phase space correction
factor $(1-p/2)$, with $p$ being the faction of 
the total baryon or antibaryon phase space that is observed, 
to the baryon or antibaryon number fluctuations in an equilibrated matter. 
We note that there is no such correction factor due to finite acceptance
if the matter is far from equilibrium when the $B\bar B$ multiplicity
distribution is Poisson.

From Eq.~(\ref{equiln}), we can also derive the following
squared net baryon number 
fluctuation for baryon species $i$ and antibaryon species $j$:
\begin{eqnarray}
\left (\Delta N_b^{ij} \right )^2 &\equiv&
\langle \left (B_i-\bar B_j \right )^2 \rangle
- \langle B_i-\bar B_j \rangle^2 \nonumber\\
&=& b_0 \langle B_i+ \bar B_j\rangle \left (1-\frac{h_i+f_j}{2} \right ). 
\end{eqnarray}
The squared net baryon fluctuation per baryon and antibaryon,
$\left (\Delta N_b \right )^2/\langle B\!+\! \bar B\rangle$, is 
thus a factor of 3 smaller in an equilibrated quark-gluon plasma 
than in an equilibrated hadronic matter, and is a useful observable as well 
\cite{Yuki,comment}. 
If the phase space acceptance is the same for baryons and antibaryons,
i.e., $h_i=f_j=p$, then the phase space correction factor is $(1-p)$
as given in Ref. \cite{Jeon}.

\section{Conclusions and Discussions}
\label{conclusion}

We have shown that $\omega_B$ or $\omega_{\bar B}$, 
the squared fluctuation of baryon or antibaryon numbers per baryon or 
antibaryon, reflects the fundamental units of baryon number  
in the matter where they are produced.
Since their expected values in the quark-gluon plasma are a 
factor of $3$ smaller than those in an equilibrated hadronic matter, 
they can be used as signatures for the quark-gluon 
plasma formed in the initial stage of relativistic heavy ion collisions.
Using a master equation with exact baryon number conservation, we have found 
that at equilibrium the baryon multiplicity distribution is not Poisson,
and its squared fluctuation is half of that expected from a Poisson 
distribution. We have also calculated the correction factors 
due to finite acceptance in experiments. Because of the contamination 
from the projectile and target nucleons, $\omega_{\bar B}$ is  
preferred over $\omega_B$. 

Although the baryon and antibaryon fluctuations are smaller in the
quark-gluon plasma formed in the initial stage of heavy ion collisions,
their magnitudes will be affected during the hadronization and subsequent
scatterings of baryons and antibaryons in the hadronic matter.
In the simplest quark coalescence model for the hadronization
of the quark-gluon plasma to the hadronic matter, where three quarks 
form a baryon and a quark-antiquark pair forms a meson, 
the baryon number fluctuation is increased as a result of the finite
fraction of quarks that form baryons. The correction factor to the baryon 
fluctuation in this hadronization scenario can be similarly derived as 
for finite acceptance. On the other hand, if hadronization is through
the formation of strings and their subsequent fragmentation \cite{lund}, 
then quark-antiquark pairs can be produced and annihilated 
during the string fragmentation and such additional processes 
usually also increase the baryon and antibaryon fluctuations. 
Furthermore, anomalous fluctuations could arise if the phase transition
is first order \cite{first1,first2} or the quark matter is at
the QCD tricritical point \cite{ebe,tri}. As to the effects of hadronic 
scatterings, elastic scatterings of baryons and antibaryons 
are not expected to affect baryon and antibaryon number fluctuations in 
the full phase space. However, baryon-antibaryon pair production 
and annihilation tend to increase the fluctuations 
towards the values expected from a hadronic matter \cite{cascade}. 
The determination of baryon and antibaryon 
fluctuations in heavy ion collisions could be further complicated by 
the existence of gluonic baryon junctions \cite{junction1,junction2,junction3},
so that baryon numbers are not necessarily carried fractionally by quarks 
but instead by the baryon junctions. In order for the baryon number 
fluctuation to be a viable signature for the quark-gluon plasma, 
quantitative studies of above effects are needed.

\section*{Acknowledgments}

We thank J. Cleymans, L.D. McLerran, K. Redlich, M. Stephanov and 
X.-N. Wang for fruitful discussions. This work was supported by 
the National Science Foundation under Grant No. PHY-9870038,
the Welch Foundation under Grant No. A-1358,
and the Texas Advanced Research Program under Grant No.
FY99-010366-0081.

{}

\begin{thebibliography}{10}
\bibitem{quark} 
The formulae in this section can be equally applied to 
quark-antiquark production, $gg \leftrightarrow q_i \bar q_j$,  
and $s$ is then the net quark number in the quark-gluon plasma.
\bibitem{Cle0} The equilibrium solution given in Eq.(\ref{equil}) 
and its first and second moments can also be derived from 
relativistic statistical mechanics as shown in Ref. \cite{Cle}.
\bibitem{comment}
The net charge and net baryon fluctuations can also be smaller than 
the values expected from a hadronic matter
if the small initial fluctuations produced from primary collisions 
could survive due to rapid expansion. See 
M. Ga\'zdzicki and S. Mr\'owczy\'nski, eprint nucl-th/0012094.
\bibitem{cascade}
Effects of partonic and hadronic scattering on the baryon number
fluctuation are being studied using transport models.
\end{thebibliography}
\end{document}